\documentclass{comad12}
\usepackage{comment}
\usepackage{graphicx}
\usepackage{epsfig}
\usepackage{algorithm} 
\usepackage{algorithmic}
\usepackage{url}
\usepackage{epstopdf}

\title{A Novel Query-Based Approach\\for Addressing Summarizability Issues in XOLAP}

\author{
	Marouane Hachicha~\hspace{2.5cm}Chantola Kit~\hspace{3cm}J{\'e}r{\^o}me Darmont\\\\
	\large{Universit{\'e} de Lyon (ERIC Lyon 2)}\\
	\large{5 avenue Pierre Mend{\`e}s-France}\\ 
	\large{69676 Bron Cedex}\\
	\large{France}\\
	\large{marouane.hachicha@univ-lyon2.fr, kchantola@gmail.com, jerome.darmont@univ-lyon2.fr}
}

\begin{document}
\maketitle

\begin{abstract}
  
The business intelligence and decision-support systems used in many application domains casually rely on data warehouses, which are decision-oriented data repositories modeled as multidimensional (MD) structures. MD structures help navigate data through hierarchical levels of detail. In many real-world situations, hierarchies in MD models are complex, which causes data aggregation issues, collectively known as the summarizability problem. This problem leads to incorrect analyses and critically affects decision making. To enforce summarizability, existing approaches alter either MD models or data, and must be applied \textit{a priori}, on a case-by-case basis, by an expert. To alter neither models nor data, a few query-time approaches have been proposed recently, but they only detect summarizability issues without solving them. Thus, we propose in this paper a novel approach that automatically detects and processes summarizability issues at query time, without requiring any particular expertise from the user. Moreover, while most existing approaches are based on the relational model, our approach focus on an XML MD model, since XML data is customarily used to represent business data and its format better copes with complex hierarchies than the relational model. Finally, our experiments show that our method is likely to scale better than a reference approach for addressing the summarizability problem in the MD context.

\end{abstract}

\section{Introduction}
\label{Introduction}

Business intelligence and decision-support systems in general are nowadays used in many business (e.g., finance, telecoms, insurance, logistics) and non-business (e.g., agriculture, medicine, health and environment) domains. Such systems casually rely on data warehouses, 
which are designed, both at the conceptual and logical levels, using multidimensional (MD) structures~\cite{RizziALT06}. In MD models, facts are analysis subjects of interest (e.g., sales) that are described by a set of (usually numerical) measures (e.g., sale quantity and amount) w.r.t. analysis axes called dimensions (e.g., book category, sale date, sale location...). Dimensions may be organized in hierarchical levels to allow data aggregation at different granularities (e.g., store, city, state or country, from the finer level to the coarser level). 

MD modeling essentially aims at easing online analytical processing (OLAP), whose main operators help navigate data through coarser (roll up) and finer (drill down) levels of detail. In this context, aggregating measures works fine when intradimensional relationships are one-to-many (e.g., a book belongs to one single category). However, in real-world situations, dimension hierarchies may be much more complex \cite{BeyerCCOPX05,Malinowski08z}, which leads to a semantic gap between MD models and current OLAP tools \cite{RizziALT06}, an issue known as the summarizability problem \cite{LenzS97}. Violating summarizability is a critical matter, for it causes erroneous aggregations and, therefore, erroneous analyses that can jeopardize important decisions \cite{MazonLT09}. However, testing summarizability is a difficult (coNP-complete) problem \cite{HurtadoGM05}. Finally, complex hierarchies are difficult to both represent in classical database management systems and query with SQL-like languages, while XML storage and interrogation with XQuery is much more natural \cite{BeyerCCOPX05}, which led to the design of XML data warehouses and so-called XOLAP solutions.
 
The summarizability problem is widely acknowledged as crucial and has received some attention in the Nineties, with most solutions aiming at \emph{a priori} normalizing data to enforce summarizability. Quite surprisingly, few researchers came back on this topic since then, although we identify two types of shortcomings in normalization approaches. 
First, normalizing data breaks initial conceptual MD models, provoking the alteration or loss of some semantics. Thus, there would be no point in exploiting XML's flexibility to model rich, complex hierarchies if they were ``flattened'' after normalization. Second, data normalization applies \textit{a priori}, on a case-by-case basis, and requires the intervention of an expert in MD modeling. Such an approach is subjective, likely to be costly and does not scale well w.r.t. data volume \cite{MazonLT08}. Finally, to the best of our knowledge, there is no existing XOLAP approach that provides a practical solution to summarizability issues, while they are much likely to occur in an XML data warehouse with complex dimension hierarchies. The closest approach does detect summarizability issues, but then returns no result \cite{PedersenPP04, PedersenRP02b}.

Thus, we propose in this paper a novel approach, set in the XOLAP context, to the summarizability problem. By contrast to normalization, our approach does not alter data to retain all semantics. We also favor paying the price of some overhead and tackle the summarizability problem at query time, without requiring any expertise beyond the user's, to avoid re-normalizing when data schema evolves, favor scalability and eliminate human-related costs. {In many institutions, decision-support applications indeed require external Web data \cite{Hackathorn99}. Due to the heterogeneity and high evolutivity of such data, an XOLAP run-time solution is more suitable than \emph{a priori} expert interventions.}

The remainder of this paper is organized as follows. In Section~\ref{sec-background}, we formalize the background information related to data warehouses, and define what we term complex hierarchies and summarizability. We also review the existing approaches for enforcing summarizability. In Section~\ref{sec-Query-based-approach}, we motivate and introduce our query-based solution to complex hierarchy management in XOLAP, including novel pattern tree-based data and query models, as well as the aggregation algorithm that exploits them. In Section~\ref{sec-experimental-validation}, we provide a complexity study and an experimental validation of our work. Finally, in Section~\ref{sec-conclusion}, we conclude this paper and hint at future research.

\section{Background}
\label{sec-background}

In this section, we formalize data warehousing concepts and define complex hierarchies that lead to summarizability issues. Then, we discuss the approaches that address the summarizability problem.

\subsection{Data Warehouses}
\label{data-warehouses}

\subsubsection{Data Warehouse}
\label{def-data-warehouse}

A data warehouse $W$ modeled w.r.t. a snowflake schema (i.e., with dimension hierarchies) is defined as $W = (\cal{F}, \cal{D})$, where
	$\cal{F}$ is a set of facts to observe and
	$\cal{D}$ is a set of dimensions or analysis axes. Let $d = |\cal{D}|$.

\subsubsection{Dimension and Hierarchy} 
\label{def-dim-and-hierarchy}

$\forall i \in [1, d]$, a dimension $D_i \in \cal{D}$ is defined as a hierarchy made up of a set of $n_i$ levels: $D_i = \{{\cal{H}}_{ij} | j = 1, n_i\}$. By convention, we denote ${\cal{H}}_{i1}$ as the lowest granularity level.
$\forall j \in [1, n_i]$, a hierarchy level ${\cal{H}}_{ij}$ is defined in intention as ${\cal{H}}_{ij} = (ID_{ij}, \{A_{ijk} | k = 1, a_{ij}\}, R_{ij})$, where
 $ID_{ij}$ is the identifier attribute of ${\cal{H}}_{ij}$,
 $\{A_{ijk}\}$ is a set of $a_{ij}$ so-called member attributes of ${\cal{H}}_{ij}$, and
 $R_{ij}$ is an attribute that references a hierarchy level at a higher granularity than that of ${\cal{H}}_{ij}$ (notion of roll up).

Let $dom()$ be a function that associates to any attribute its definition domain. Let $h_{ij} = |{\cal{H}}_{ij}|$. $\forall l \in [1, h_{ij}]$, instances
of ${\cal{H}}_{ij}$ are tuples $H_{ijl} = (\sigma_{ijl}, \{\alpha_{ijkl} | k = 1, a_{ij}\}, \rho_{ijl})$, where
$\sigma_{ijl} \in dom(ID_{ij})$,
$\alpha_{ijkl} \in dom(A_{ijk})$ $\forall k \in [1, a_{ij}]$, and
$\rho_{ijl} \in dom(ID_{ij'})$ with $j' \in [1, n_i]$.

\subsubsection{Fact}
\label{Def-Fact}

The set of facts~$\cal{F}$ is defined in intention as $\cal{F} =$ $(\{\Delta_i | i = 1, d\}, \{M_j | j = 1, m\})$, where $\{\Delta_i\}$ is a set of $d$ attributes that reference instances of hierarchy levels ${\cal{H}}_{i1}$ of each dimension $D_i \in \cal{D}$, and $\{M_j\}$ is a set of $m$ measure (or indicator) attributes that characterize facts.

Let $f = |\cal{F}|$. $\forall k \in [1, f]$, instances of $\cal{F}$ are tuples $F_k = (\{\delta_{ik} | i = 1, d\}, \{\mu_{jk} | j = 1, m\})$, where
$\delta_{ik} \in dom(ID_{i1})$ $\forall i \in [1, d]$, and 
$\mu_{jk} \in dom(M_j)$ $\forall j \in [1, m]$.

\subsection{Complex Hierarchies}
\label{Def-Complex-Hierarchy}

We term a dimension hierarchy $D_i$ as complex if it is both non-strict and incomplete. We choose this new, general denomination because dimension hierarchy characterizations vary wildly in the literature. For example, Beyer et al. name complex hierarchies ragged hierarchies \cite{BeyerCCOPX05}, while Rizzi defines ragged hierarchies as incomplete only \cite{Riz07}. Malinowski and Zim{\'a}nyi also use the terms of complex generalized hierarchy \cite{Malinowski08z}, but even though they include incomplete hierarchies, they do not include non-strict hierarchies.

\subsubsection{Non-Strict Hierarchy}
\label{sec:HierarchieNonStricte}

A hierarchy is non-strict \cite{AbelloSS06,MalinowskiZ06,Tor03} or multiple-arc \cite{Riz07} when attribute $R_{ij}$ is multivalued. In other terms, from a conceptual point of view, a hierarchy is non-strict if the relationship between two hierarchical levels is many-to-many instead of one-to-many. For example, in a dimension describing products, a product may belong to several categories instead of just one.

Similarly, a many-to-many relationship between facts and dimension instances may exist \cite{Riz07}. For instance, in a sale data warehouse, a fact may be related to a combination of promotional offers rather than just one. Formally, here, attributes $\Delta_i$ $(\forall i \in [1, d])$ may be multivalued. 

\subsubsection{Incomplete Hierarchy}
\label{sec:HierarchieNonCouvrante}

A hierarchy is incomplete \cite{DyresonPJ03,PourabbasR00}, non-covering \cite{AbelloSS06,MalinowskiZ06,Tor03} or ragged \cite{Riz07} if attribute $R_{ij}$ allows linking a hierarchy level ${\cal{H}}_{ij}$ to another hierarchy level ${\cal{H}}_{ij'}$ by ``skipping'' one or more intermediary levels, i.e., $R_{ij}$ refers to $ID_{ij'}$ such that $j' > j + 1$.
This occurs, for instance, if in a dimension describing stores, the store-city-state-country hierarchy allows a store to be located in a given region without being related to a city (stores in rural areas). 

Similarly, facts may be described at heterogeneous granularity levels. For example, still in our sale data warehouse, sale volume may be known at the store level in one part of the world (e.g., Europe), but only at a more aggregate level (e.g., country) in other geographical areas. This means that $\forall i \in [1, d]$, $\delta_i \in dom(ID_{ij})$ with $j \in [1, n_i]$ (constraint $j = 1$ is forsaken).

A particular case of incomplete hierarchies are called non-onto \cite{PedersenJD99}, heterogeneous \cite{HurtadoGM05}, unbalanced \cite{HummerLBS02,Malinowski08z} or asymmetric \cite{MalinowskiZ06} hierarchies. A hierarchy is non-onto when all paths from the root to a leaf in the hierarchy do not have equal lengths \cite{PedersenJD99}, but here, missing elements are always child nodes, while they may be parent nodes in an incomplete hierarchy.

\vspace{0.3cm} Note that some papers addressing the summarizability problem differentiate between intradimensional relationships and fact-to-dimension relationships \cite{MazonLT08}. By contrast, as Pedersen et al. \cite{PedersenJD99}, we consider that summarizability issues and solutions are the same in both cases, since facts may be viewed as the very finer granularity in the dimension set.

\subsection{Summarizability in MD Models} 
\label{sec-summarizability}

The notion of summarizability was introduced by Rafanelli and Shoshani in the context of statistical databases \cite{RafanelliS90}, where it refers to the correct computation of aggregate values with a coarser level of detail from aggregate values with a finer level of detail. Then, Lenz and Shoshani defined three constraints that guarantee summarizability in the MD context \cite{LenzS97}: (1) hierarchies must be strict; (2) hierarchies must be complete; (3) aggregate data types must be compatible, i.e., an aggregate function must be applicable to a given measure for a given set of dimensions. For instance, a maximum sale amount is a meaningful aggregation, while a sum of temperatures would be meaningless. These constraints also hold for fact-to-dimension relationships \cite{MazonLT08}. In this paper, we assume that the type compatibility constraint is handled by users.


One way to ensure summarizability in a MD model is to simply disallow complex hierarchies at design time, as in the Dimensional Fact Model \cite{GolfarelliMR98}. However, to support different kinds of complex real-world situations, most MD models do allow complex hierarchies. Thence, the summarizability problem must be addressed. There are two main families of approaches: schema normalization and data transformation, which are reviewed below. Both families of approaches operate at design time. 

More recent proposals operate at query time, but they are very few. Guidelines have been proposed for tolerating and displaying incorrect aggregation results \cite{HornerS05}, but they have not been implemented. The generalized projection XOLAP operator \cite{PedersenPP04, PedersenRP02b} detects summarizability issues, but does not solve them and returns an error flag instead.

Finally, the interested reader may find more details about summarizability issues in the survey by Maz{\'o}n et al. \cite{MazonLT09}.

\subsubsection{Schema Normalization}
\label{sec-Normalization-approaches}

Two strategies may be used to achieve schema normalization. The first strategy is based on the definition of constraints and transformation rules. For instance, Hurtado et al. propose a class of integrity constraints to address incompleteness, namely dimension constraints and frozen dimensions \cite{HurtadoGM05}. Frozen dimensions are minimal, complete dimensions mixed up in incomplete dimensions using dimension constraints that help model incomplete hierarchy schemas. From their part, Lechtenb{\"o}rger and Vossen introduce new MD normal forms (MNFs)~\cite{LechtenborgerV03}. 1MNF does not allow non-strict hierarchies, while 2MNF and 3MNF permit to model incomplete relationships using context dependencies, i.e., dimension constraints. Specialization constructs in dimensions can lead to incomplete relationships~\cite{LenzS97,RafanelliS90} and context dependencies enable an implicit representation of such specializations.

The second strategy adds new structures into the model in order to ensure summarizability. In relational implementations, bridge tables are used to capture non-strict fact-to-dimension relationships via foreign keys that refer to the dimension and fact tables \cite{kimball02,SongRME01}. 
Arguing that bridge tables defined at the logical level make the modeling of complex structures difficult, some authors introduce their equivalent at the conceptual level \cite{MalinowskiZ06,MazonLT08}. Such additional entities/classes help store instances at the origin of incompleteness and/or non-strictness. Finally, Mansmann and Scholl propose a two-phase modeling approach that transform incomplete hierarchies into a set of well-behaved sub-hierarchies without summarizability problems~\cite{MansmannS06, MansmannS07}.

\subsubsection{Data Transformation}
\label{Data-transformation-approaches}

The reference data transformation approach by Pedersen et al. transforms dimension and fact instances to enforce summarizability \cite{PedersenJD99}. To solve incompleteness, all mappings between hierarchical levels are transformed to be complete with the help of an algorithm named \texttt{MakeCovering}. For example, suppose that some addresses are missing in an address-city-country hierarchy. \texttt{MakeCovering} inserts new values into the missing hierarchical level address to ensure that mappings to higher hierarchical levels are summarizable. \texttt{MakeCovering} exploits metadata and/or expert advice for this sake. For example, an expert would be required to recover missing addresses in small streets in the USA or Australia. The authors also propose a simplified version of \texttt{MakeCovering}, \texttt{MakeOnto}, to handle summarizability in non-onto hierarchies by replacing childless nodes by so-called placeholder values.

Mappings are made strict with the help of another algorithm named \texttt{MakeStrict}. \texttt{MakeStrict} avoids ``double counting'' by ``fusing'' multiple values in a parent hierarchical level into one ``fused'' value, and then linking the child value to the fused value. Fused values are inserted into a new hierarchical level in-between the child and the parent. Reusing this new level for computing higher-level aggregate values leads to correct aggregation results. 

Mansmann and Scholl further modify Pedersen et al.'s algorithms to eliminate roll up/drill down incomplete and non-strict hierarchies at the instance level \cite{MansmannS06, MansmannS07}. Finally, Li et al. demonstrate that \texttt{MakeCovering} does not work on some real-world cases, i.e., geographical hierarchies in China \cite{LiSZY05}. They identify four types of incompleteness that are specific to China and thence propose several variations of \texttt{MakeCovering} to handle them.

\section{Query-based Complex Hierarchy Management in XOLAP} 
\label{sec-Query-based-approach}

\subsection{Motivation and Contributions}
\label{sec:MotivationAndContributions}

In XML data warehouses and XOLAP, complex data structures, and especially complex hierarchies, are likely to be present, and are likely to evolve with time faster than in legacy decision-support systems. In such a context, summarizability cannot be enforced through a costly \cite{MazonLT08} data normalization process each time schema and data are updated. Thus, as in the most recent existing approaches \cite{HornerS05,PedersenPP04,PedersenRP02b}, we advocate for a run-time solution.

However, while existing run-time approaches do detect summarizability issues and warn the user, they still output incorrect or absent results. Our first contribution is thus to complete the process and output correct results. To achieve this goal, we adapt and automatize well-known solutions from the literature (Sections \ref{sec-principle} and \ref{sec-Algorithms-for-query-time-summarizability}). Since we operate at query time, we deliberately adopt simple and robust solutions not to add too much overhead over summarizability testing. Such reference approaches are still customarily reused and adapted by recent approaches \cite{MansmannS06, MansmannS07}.

Furthermore, all XOLAP approaches we are aware of propose operators under the form of ad-hoc programs, and rely on relational database systems, including Pedersen et \emph{al.}'s \cite{PedersenPP04, PedersenRP02b}. By contrast, we aim at contributing to build an XOLAP algebra that can later translate into standard XQuery statements. Thus, our second contribution introduces data and query models based on the data trees and tree patterns used in XML processing~\cite{HachichaD-tkde12}, respectively (Section \ref{sec-dataModel}).

\subsection{Principle of our Approach}
\label{sec-principle}

\begin{figure}[hbt]
\centering
\epsfig{file=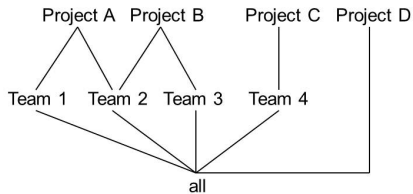, width=6cm}
\caption{Sample complex hierarchy}
\label{complexH}
\end{figure}

To illustrate how our approach operates, let us consider the example from Figure~\ref{complexH}, which represents a complex ``project management'' hierarchy at the instance level, adapted from \cite{MansmannS06, MansmannS07}. This hierarchy is non-strict because teams may manage several projects (Team 2 manages projects A and B), while projects may be managed by several teams (projects A and B are managed by teams 1 and 2, and teams 2 and 3, respectively). The hierarchy is also incomplete, since Project D is not managed by any particular team; thus it is complex. 

First, to handle non-strict hierarchies in a given dimension $D_i$, we must avoid multiplying the aggregation of instance measures of a hierarchy level ${\cal{H}}_{ij}$ when rolling up to level ${\cal{H}}_{ij+1}$. Thus, when building the set of groups $G$ with respect to a grouping criterion, we fuse multiple values in ${\cal{H}}_{ij+1}$ into one single ``fused value'', i.e., we build $G = \bigcup\limits_{l \in [1, h_{ij}]} \rho_{ijl}$, where multivariate values of $\rho_{ijl}$ are considered as sets instead of single values. In our example, suppose we are counting projects per teams for projects A and B. Then $G_{NS} = \{\{Team~1, Team~2\}, \{Team~2, Team~3\}\}$. The number of projects in $\{Team~1, Team~2\}$ is 1, the number of projects in $\{Team~2, Team~3\}$ is 1, for a correct total of 2. If $G_{NS}$ had been $\{Team~1, Team~2, Team~3\}$, the total number of projects would have been wrong ($1 + 2 + 1 = 4$) in ${\cal{H}}_{12}$. 

Second, to handle incomplete hierarchies, we must, when rolling up from a hierarchy level ${\cal{H}}_{ij}$ to level ${\cal{H}}_{ij+1}$, still aggregate measures of instances of ${\cal{H}}_{ij}$ that are \emph{not} present in ${\cal{H}}_{ij+1}$. Thus, when building $G$, all ``missing instances'' are grouped into an artificial ``Other'' group, i.e., $G = \bigcup\limits_{l \in [1, h_{ij}]} \rho_{ijl}$ $\cup \{Other\}$ such that $\exists l' / \rho_{ijl} = \sigma_{i(j+1)l'}$. In our example, suppose we are again counting projects per teams, but for projects C and D. Then $G_I = \{Team~4, Other\}$. The number of projects in $G_I$ is 2, whereas it would have been wrong, i.e., 1, if $G_I$ had been $\{Team~4\}$ only.

Third, to handle complex hierarchies, we simply apply both the managements of non-strict and incomplete hierarchies. Thus, here, $G = \bigcup\limits_{l \in [1, h_{ij}]} \rho_{ijl} \cup \{Other\}$ such that $\exists L / \rho_{ijl} = \bigcup\limits_{l' \in L} \sigma_{i(j+1)l'}$. In our example, if we are now counting projects per teams for all projects, then $G_C = G_{NS} \cup G_I$, and the number of projects in $G_C$ is correct, i.e., 4.

		Finally, note that, beyond the expert-based preprocessing vs. our automatic, on-the-fly approach, there is a substantial difference between our view of incomplete hierarchy management and Pedersen et \emph{al.}'s reference solution \cite{PedersenJD99}. While they call to an expert to replace all ``missing values'' in $G$ by actual values, we indeed automatically add an ``Other'' group for all ``missing values'' of a given hierarchical level. ``Other" values from different hierarchy levels are of course distinguished, e.g., \texttt{Project[Other]} is different from \texttt{Team[Other]}.

		Thus, we presumably loose in semantical finesse, but we spare the cost of the expert. Moreover, the simplicity of our approach helps handle all cases of incompleteness identified by Li et al. \cite{LiSZY05}, while \texttt{MakeCovering} cannot.

\subsection{Data and Query Models} 
\label{sec-dataModel}

\subsubsection{Data Model}
\label{sec:DataModel}

Since complex hierarchies have been shown to be better represented in XML at the physical level \cite{BeyerCCOPX05}, we choose XML to model MD data. Thus, at the logical level, we choose XML data trees to model MD structures. Data trees are indeed casually used to represent and manipulate XML documents, whose hierarchical structure is akin to a labeled ordered, rooted tree \cite{HachichaD-tkde12}. Moreover, data trees allow modeling MD structures. Formally, a data tree $t$ models an XML document or a document fragment. It can be defined as a triple $t = \left(r, N, E\right)$, where $N$ is the set of nodes, $r \in N$ is the root of $t$, and $E$ is the set of edges stitching together couples of nodes $(n_{i}, n_{j}) \in N \times N$. 

\begin{figure} [hbt]
\centering
\epsfig{file=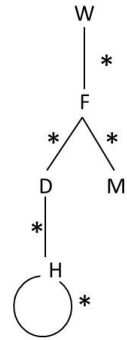, width= 2 cm}
\caption{Multidimensional data tree model}
\label{datamodel}
\end{figure}

Figure~\ref{datamodel} shows how we logically model MD data with a data tree. $\ast$-labeled edges indicate a one-to-many relationship. The data tree root, $W$, models the data warehouse. Its child nodes $F$ model facts. Each fact is described by a set of dimensions $D$ and measures $M$. For a given fact, we may have several dimensions (such as client, supplier...) and several measures (such as account, quantity...). A dimension hierarchy can have any number of levels $H$. The $\ast$ multiplicity on the $D$-$H$ edge allows facts to roll up to any number of hierarchy levels, at any granularity (fact-to-dimension relationships). The recursive edge on $H$ allows any hierarchy level to roll up to several higher levels, possibly skipping any number of intermediary levels (intradimension relationships). Thus, this representation permits to model complex hierarchies.

\begin{figure*} [hbt]
\centering
\epsfig{file=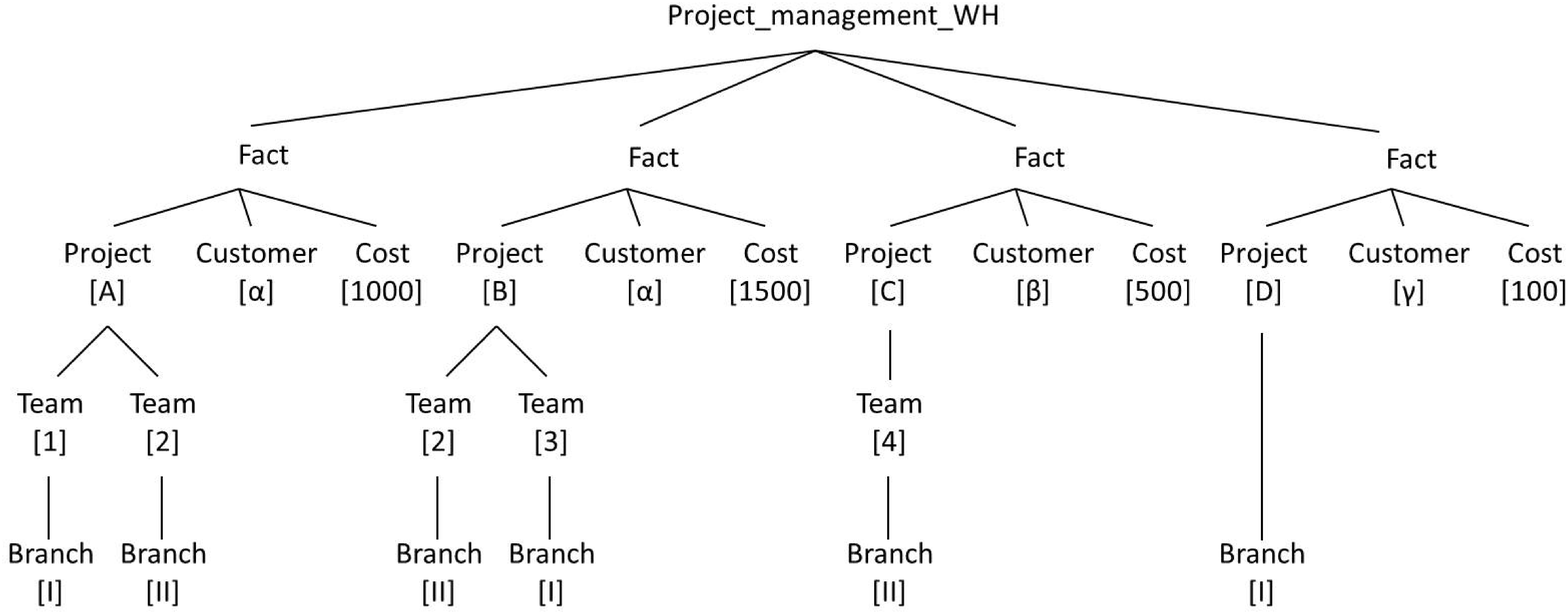, width=13 cm}
\caption{Sample multidimensional data tree}
\label{datatree}
\end{figure*}

Figure~\ref{datatree} exemplifies the instantiation of our model by elaborating on the complex hierarchy from Figure~\ref{complexH}. Here, facts are described by a project and a customer dimension, and the only measure is cost.
                      
\subsubsection{Query Model}
\label{sec:QueryModel}

\begin{figure} [hbt]
\centering
\epsfig{file=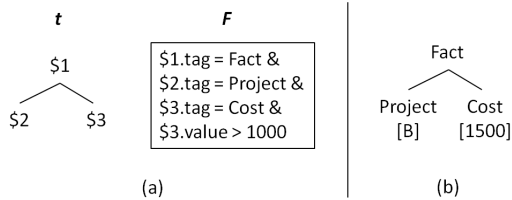, width=7.5cm}
\caption{Sample pattern (a) and witness (b) trees}
\label{ptwt}
\end{figure}

Since we use XML data trees as our logical data model, we use XML tree patterns, which are the most efficient structures to query data trees \cite{HachichaD-tkde12}, as our query model. A tree pattern (TP) or tree pattern query is a pair $\left(t, F\right)$ where $t$ is a data tree $(r, N, E)$. An edge in $t$ may either be a parent-child (\textit{pc} for short, simple edge in XPath) node relationship or an ancestor-descendant (\textit{ad} for short, double edge in XPath) node relationship. $F$ is a formula that specifies constraints on TP nodes. More explicitly, $F$ is a boolean combination of predicates on TP node values. For example, the TP from Figure~\ref{ptwt}(a) selects all projects whose cost is strictly greater than 1000. Matching this TP against the data tree from Figure~\ref{datatree} outputs a new data tree, also called witness tree (WT), which is depicted in Figure~\ref{ptwt}(b). 

\begin{figure} [hbt]
\centering
\epsfig{file=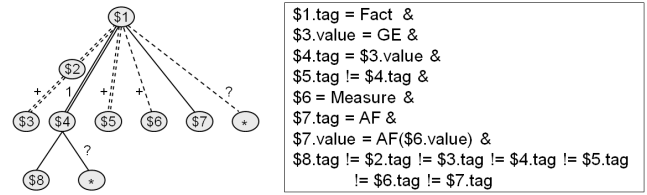, width=8.5cm}
\caption{Multidimensional tree pattern model}
\label{tpmodel}
\end{figure}

To help query MD data modeled w.r.t. Figure~\ref{datamodel}'s data tree model, we propose the TP model depicted in Figure~\ref{tpmodel}. In this TP model, nodes connected to their parent nodes with a dotted edge do not appear in the WT, unlike nodes connected to their parent nodes with a solid edge. Moreover, for each edge ($u$, $v$) where $u$ is a parent (or an ancestor) of $v$: a ``+'' label means that one or many matches of $v$ are allowed for each match of $u$ in a WT; a ``?'' label means that zero to one match of $v$ is allowed for each match of $u$ in a WT; and a ``1'' label means that one and only one match of $v$ is allowed for each match of $u$ in a WT. Nodes from our TP model are tagged with \$$i$ ($i$ being a number) or with~$\ast$. Nodes tagged with $\ast$ are always connected to their parent nodes with a $pc$ relationship (/). In XPath~2.0~\cite{BerglundBCFKRS10}, a path $x$/$\ast$ such that $x$ is a node returns a different result from the path $x$//$\ast$.~$x$/$\ast$ returns the hierarchy connected to $x$ while $x$//$\ast$ returns the same result as $x$/$\ast$ but with duplicate nodes. Thus, we choose to respect the XPath 2.0 standard.

Formula $F$ precises how our TP model matches a MD data tree, as follows. Node~\$1 matches one fact. Node~\$3 specifies one to many grouping elements (denoted GE in $F$). A grouping element \$3 is a hierarchical level. Node~\$2 models all nodes that may exist between~\$1 and~\$3. Node~\$4 receives~\$3's content in order to match only one node corresponding to each grouping element. Node~\$4's content finally receives a group $G$ (Section~\ref{sec-principle}). Node~\$5~matches all dimension hierarchical levels different from \$4. These matched nodes do not appear in the WT because the corresponding dimensions do not belong to grouping elements. \$6 specifies one to several measures required for aggregation purposes. Node~\$7 stores the result of applying an aggregation function (e.g., sum, count, etc.) AF on nodes~\$6. There is no guarantee that all the nodes output when matching the $\ast$ child nodes of \$4 against a MD data tree appear in the WT due to incomplete hierarchies. Thus, node \$8 retains the matching result of the $\ast$ child nodes of \$1, except measures not used in any aggregation.

\subsection{Grouping and Roll up Algorithms}
\label{sec-Algorithms-for-query-time-summarizability}

In this section, we translate the principles from Section~\ref{sec-principle} into a grouping algorithm called Query-Based Summarizability (\texttt{QBS}) that exploits the data and query models from Section~\ref{sec-dataModel}. Then, we devise a roll up operator based on \texttt{QBS}. 

\texttt{QBS} (Algorithm~\ref{roll-up-2}) essentially processes a ``group by'' query with respect to any number of grouping criteria, and additionally handles summarizability issues on the fly. \texttt{QBS} inputs: (1) a data tree $D$ modeled w.r.t. Figure~\ref{datamodel}'s data tree model and (2) a TP $TPQ$ modeled w.r.t. Figure~\ref{tpmodel}'s TP model. \texttt{QBS} outputs a list of WTs $WTlist$ (i.e., a set of at least one WT). \texttt{QBS} proceeds into two main steps: (1) incompleteness and non-strictness management; (2) group matching to construct correct aggregation results.

\begin{algorithm}[hbt]
\caption{QBS grouping algorithm}
\label{roll-up-2}
\begin{algorithmic}[1] 
\begin{small}
\STATE{\texttt{Input:}}
\STATE{~~~~~$D$~~~~~~//~Data tree}
\STATE{~~~~~$TPQ$~//~Tree pattern}
\STATE{$WTlist \leftarrow \emptyset$}
\FORALL{\$1}
	\STATE{// Step \#1: Summarizability processing}
	\STATE{$Group\_list$ $\leftarrow \emptyset$}    
	\FORALL{\$4} 
		\STATE{$Group$ $\leftarrow$ $Group~\cup$ \$4.value} 
		\IF{\$4 $\notin$ \$1.\texttt{children()} and \textit{Group}.\texttt{nbElements()} $<$ \$1.\texttt{currentChild()}.\texttt{nbChildren()}}
   		\STATE{$Group$ $\leftarrow$ $Group~\cup$ ``Other''} 
		\ENDIF  
		\STATE{$Group\_list \leftarrow Group\_list \cup Group$}  
	\ENDFOR 
	\STATE{// Step \#2: Group matching}
	\STATE{$WT \leftarrow WTlist$.\texttt{exists(}$Group\_list$\texttt{)}}
	\IF{$WT \neq \emptyset$}
  	\STATE{$WT$.\texttt{update(}\$6, \$7\texttt{)}}
  \ELSE
  	\STATE{$WT$.\texttt{create(}$D$, $TPQ$\texttt{)}} 
  	\STATE{$WTlist \leftarrow WTlist \cup WT$}
	\ENDIF
\ENDFOR 
\RETURN{\texttt{product(}$WTlist$\texttt{)}}
\end{small}
\end{algorithmic}
\end{algorithm}

More precisely, \texttt{QBS} first initializes $WTlist$ to empty. Then, for each fact, a variable $Group\_list$, which stores together all possible groups from different grouping elements, is also initialized to empty. Such groups are stored in the $Group$ variable, which comprises node values matched by \$4 in $TPQ$ (Step \#1). In case of missing instances from a hierarchical level of the grouping element (if statement on line 10), the ``Other'' value is concatenated to $Group$. The test on line 10 means that \$4 is not a child (i.e., dimension) node of the current fact and the number of elements in \textit{Group} is inferior to the number of edges rooted at the current dimension node (i.e., presence of an incomplete hierarchy). When a new group list is about to be built, the algorithm tests its existence in $WTlist$, i.e., it tests whether there exists a WT from $WTlist$ where a node tagged with the same grouping elements has a value equal to the group list's. If true, the aggregation node is updated with current measures. Otherwise, a new WT is added to $WTlist$ w.r.t. $TPQ$. Finally, all WTs are regrouped together under a unique root with the help of the \texttt{product()} function.

The description of all functions called in \texttt{QBS} follows.

\begin{itemize}

\item  $x$.\texttt{children()} returns the set of child nodes of node $x$.

\item  $x$.\texttt{nbChildren()} returns the number of children of node $x$. If our context, this function returns the number of edges rooted at $x$. 

\item $x$.\texttt{currentChild()} returns the current child of node $x$.

\item  $G$.\texttt{nbElements()} computes the number of elements in group $G$. 

\item $Tlist$.\texttt{exists(}$Glist$\texttt{)} returns the data tree containing group $Glist$ from one of the trees of $Tlist$, and $\emptyset$ otherwise.

\item  $T$.\texttt{update(}$x$, $y$\texttt{)} updates the value of node $y$ from tree $T$ with the value of node $x$. 

\item  $T$.\texttt{create(}$D$, $TPQ$\texttt{)} creates a tree $T$ by matching TP $TPQ$ against data tree $D$. 

\item \texttt{product(}$Tlist$\texttt{)} regroups together all trees from tree set $Tlist$ under one single root.
\end{itemize}

\begin{figure*} [hbt]
\centering
\epsfig{file=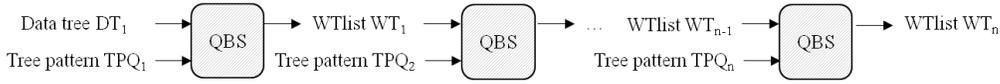, width=14cm}
\caption{Roll up process}
\label{roll-up-process}
\end{figure*}

\begin{figure} [hbt]
\centering
\epsfig{file=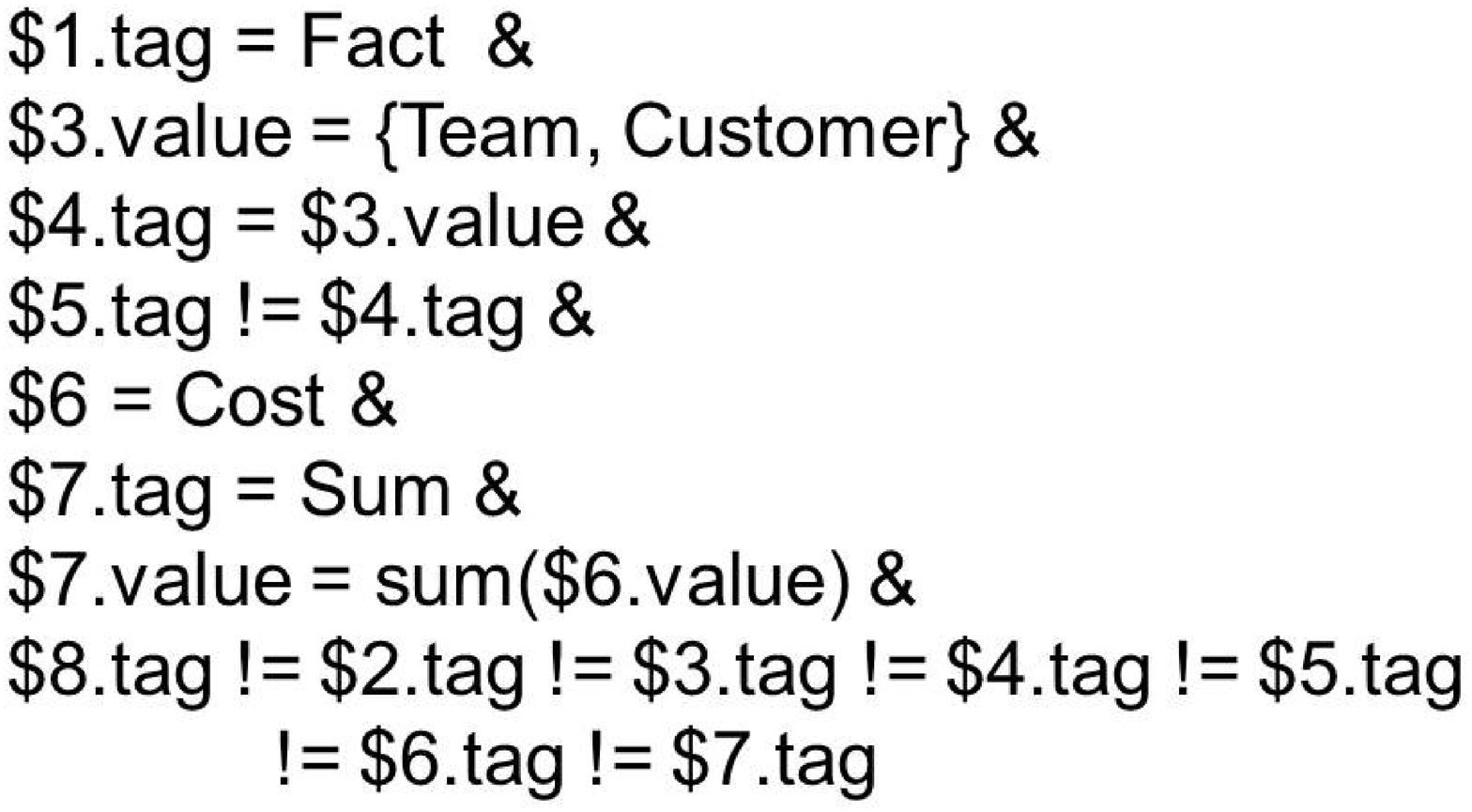, width=6 cm}
\caption{Q1 TP formula}
\label{Q1-rollup-tps}
\end{figure}

Eventually, a roll up operation is simply achieved by calling \texttt{QBS} several times, in sequence, with the output tree of each stage becoming the input tree of the next stage (Figure~\ref{roll-up-process}). For example, let us consider the MD data tree from Figure~\ref{datatree} and query Q1 = ``total cost of projects per team and per customer'', which translates into a TP whose formula is provided in Figure~\ref{Q1-rollup-tps}.

For fact Project[A], \texttt{QBS} builds $Group =$ 1-2. A first WT is thus created into $WTlist$ w.r.t. Figure~\ref{Q1-rollup-tps}'s TP, with dimension nodes (grouping element instances) Team[1-2] and Customer[$\alpha$], and an aggregation node Sum[1000]. For fact Project[B], $Group =$ 2-3 is built. Then, the algorithm checks whether there exists a WT in $WTList$ containing the $Group\_list$ (Team[2-3], Customer[$\alpha$]). As the answer is no, a second WT is created with dimension nodes Team[2-3] and Customer[$\alpha$], and aggregation node Sum[1500]. Similarly, for fact Project[C], $Group =$ 4 is built and a new WT is created with dimension nodes Team[4] and Customer[$\beta$], and aggregation node Sum[500]. 

\begin{figure*} [hbt]
\centering
\epsfig{file=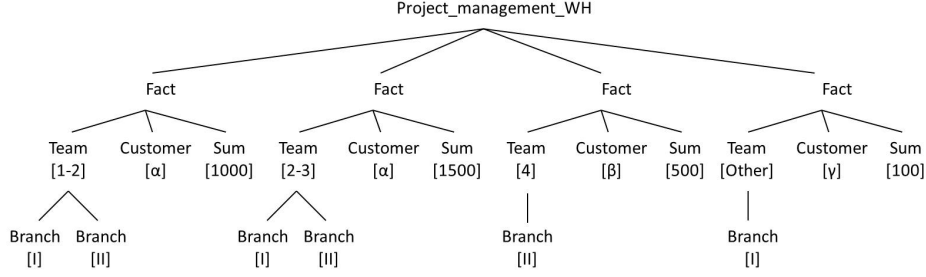, width= 13 cm}
\caption{Sample roll up operation -- Step \#1}
\label{rollup1-result}
\end{figure*}

For fact Project[D], there is no grouping element. Thus, we build $Group =$ Other and a new WT is created with dimension nodes Team[Other] and Customer[$\gamma$], and aggregation node Sum[100]. Here, \texttt{QBS} traverses all elements of the hierarchy associated to Project[D] before assigning ``Other'' to $Group$. Finally, all created WTs in $WTlist$ are appended under the same root (Figure~\ref{rollup1-result}). Note that the hierarchy of branches is always saved in WTs. \texttt{QBS} exploits the hierarchy schema (metadata) to consider Group[Other] as the parent element of Branch[I] in the corresponding WT.

\begin{figure} [hbt]
\centering
\epsfig{file=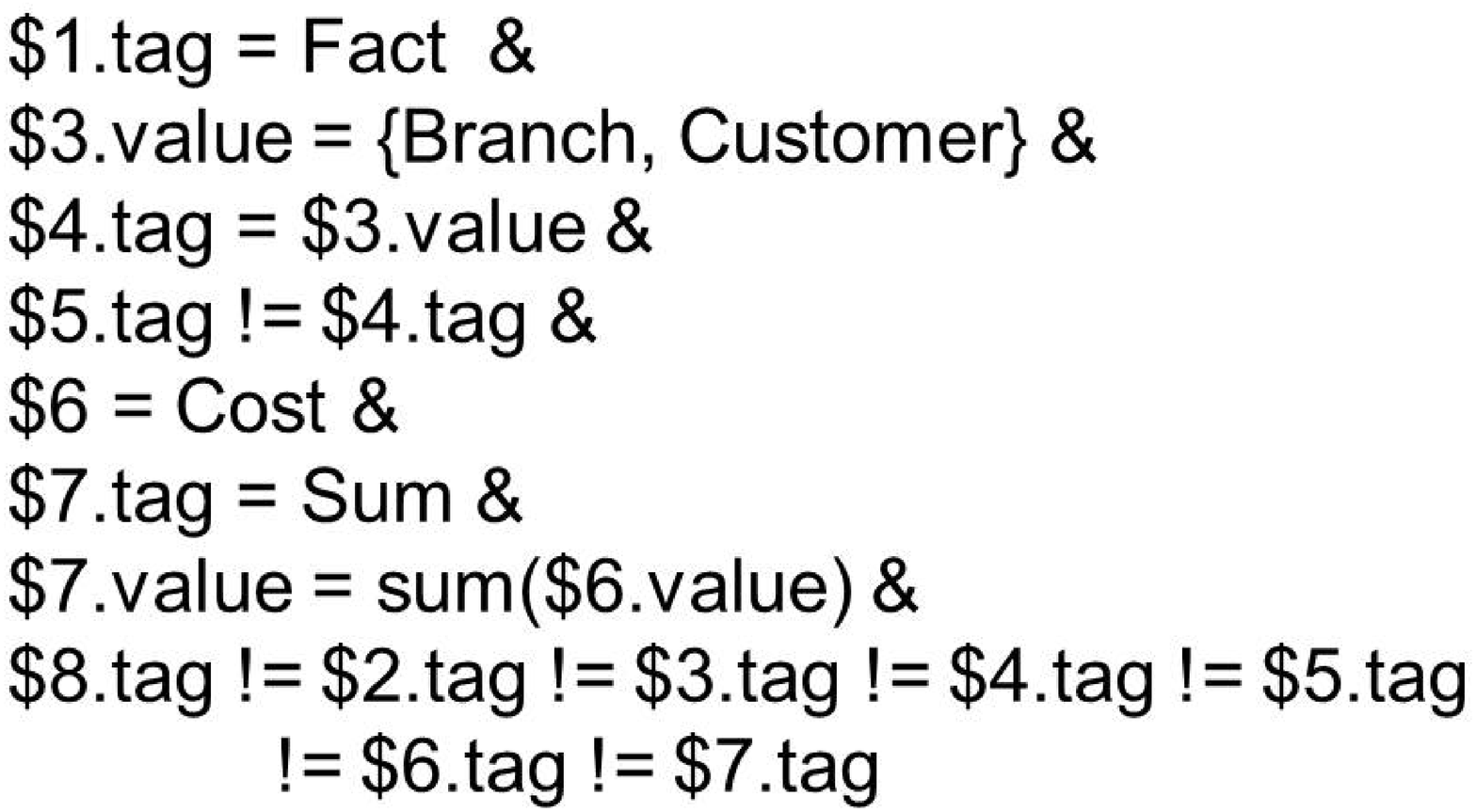, width=6 cm}
\caption{Q2 TP formula}
\label{Q2-rollup-tps}
\end{figure}

To complete the roll up operation, i.e., aggregating on branches from the aggregation already computed on groups, \texttt{QBS} inputs a new TP corresponding to Q2 = ``total cost of projects per branch and per customer'', whose formula is given in Figure~\ref{Q2-rollup-tps}, and the result tree from Figure~\ref{rollup1-result}.

\begin{figure*} [hbt]
\centering
\epsfig{file=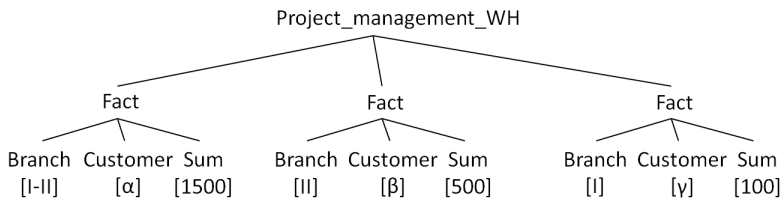, width= 11 cm}
\caption{Sample roll up operation -- Step \#2}
\label{rollup2-result}
\end{figure*}

For fact Team[1-2], $Group = $ I-II is built and a WT is created with dimension nodes Branch[I-II] and Customer[$\alpha$], and aggregation node Sum[1000]. For fact Team[2-3], $Group$ $=$ I-II is built. Then, \texttt{QBS} checks whether $Group\_list$ (Branch [I-II], Customer[$\alpha$]) exists in $WTlist$. Here, the answer is yes, and the value of the Sum node in the returned WT is updated to 2500. For fact Group[4], $Group =$ II is built. After checking the presence of $Group\_list$ (Branch[II], Customer[$\beta$]) in $WTlist$ (which is negative), a new WT is created with dimension nodes Branch[II] and Customer[$\beta$], and aggregation node Sum[500]. For fact Team[Other], a new WT is created with dimension nodes Branch[I] and Customer[$\gamma$], and aggregation node Sum[100]. Finally, all created WTs in $WTList$ are again appended under the same root (Figure~\ref{rollup2-result}).  

\section{Validation}
\label{sec-experimental-validation}

Although we should test our approach against other query-based \cite{HornerS05}, and especially XOLAP \cite{PedersenPP04, PedersenRP02b} approaches, these approaches do detect summarizability issues, but then do not output actual aggregates. Thus, though Pedersen et al.'s reference approach \cite{PedersenJD99} and its fairly recent enhancements \cite{MansmannS06, MansmannS07} apply once \emph{a priori}, we can only compare our approach with it. Moreover, we particularly focus on the \texttt{MakeStrict} and \texttt{MakeCovering} algorithms, since \texttt{MakeCovering} generalizes \texttt{MakeOnto}. For conciseness, we label the combination of \texttt{MakeStrict} and \texttt{MakeCovering} as \texttt{Pedersen} in the following.

\subsection{Complexity Study}
\label{sec-time-complexity}

Let us recall Section~\ref{data-warehouses}'s notations: $f$ is the number of facts in the data warehouse and $d$ the number of dimensions. Moreover, let $s$ be number of subdimensions, i.e., branches in non-strict hierarchies. When processing summarizability in \texttt{QBS} (Step \#1 of the algorithm), for each fact and each dimension, we need to check missing values in hierarchies and replace them by ``Other'', and then to check whether the value exists in the current group. Thus, $f \times d \times (1 + 2 + ... + s-1)$ tests must be performed in the worst case. Thus, the complexity of summarizability processing is $\cal{O}$$(fds^2)$. 

Furthermore, when performing aggregation, for each fact, we need to check whether a group exists. Following the same reasoning, the complexity of group matching (Step \#2 of \texttt{QBS}) is thus $\cal{O}$$(f^2ds^2)$. Thus, the global complexity of \texttt{QBS} is $\cal{O}$$(f^2ds^2) + \cal{O}$$(fds^2) = \cal{O}$$(f^2ds^2) \approx \cal{O}$$((fds)^2)$. 

Since $fds$ represents the input size, if we state that $n = fds$, then the worst-case complexity of \texttt{QBS} is $\cal{O}$$(n^2)$, i.e., the same as \texttt{Pedersen}'s \cite{PedersenJD99}. The worst case occurs when using linear search in the algorithms. Using binary search instead should bring complexity down to $\cal{O}$$(n$ log $n)$ in most realistic scenarios \cite{PedersenJD99}.

\subsection{Experimental Validation}

\subsubsection{Experimental Setup}
\label{subsec-experimental-setup}

To compare \texttt{QBS} to \texttt{Pedersen}, we use the XWeB benchmark \cite{xweb-tpctc10}, which remodels the TPC-H \cite{tpch} relational database as a star XML schema. 
XWeB initially generates documents scaling in size from 50,000 to 250,000 facts. The first and second rows of Table~\ref{table-data-size} range generated data in number of facts and kilobytes (mininum size is 13~MB and maximum size 67~MB). 

\begin{table*}
\centering
\caption{Dataset size (KB)}
\label{table-data-size}
{\small
\begin{tabular}{|l|r|r|r|r|r|r|r|r|} 
\hline
No. Facts&50,000&100,000&150,000&200,000&250,000\\ 
\hline
XWeB&13,661&27,366&41,070&54,775&68,479\\ \hline
XWeB DT&27,700&55,390&82,800&110,577&138,015\\ \hline
Incomplete 5\%&27,626&55,242&82,543&110,249&137,573\\ \hline
Non-strict 5\%&28,669&57,328&85,671&114,422&142,786\\ \hline
Complex 5\%&28,376&56,742&84,791&113,252&141,319\\ \hline
Incomplete 50\%&25,020&50,030&74,769&99,842&124,601\\ \hline
Non-strict 50\%&35,412&70,826&105,914&141,397&176,527\\ \hline
Complex 50\%&32,522&65,031&97,263&129,839&162,088\\ \hline
\end{tabular}}
\end{table*}

\begin{figure}[hbt]
\centering
\epsfig{file=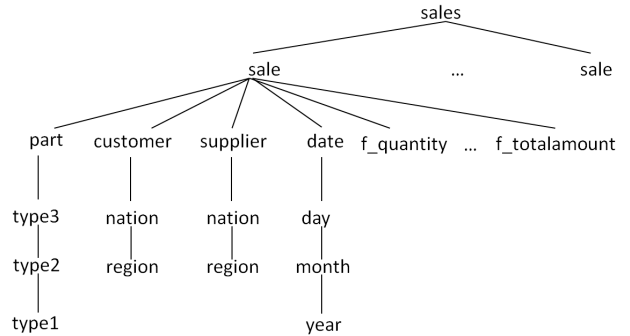, width=8.5cm}
\caption{XWeB data tree model}
\label{fig-xweb-data-tree-model}
\end{figure}

Then, since XWeB's data warehouse is not modeled w.r.t. our data tree model (Section~\ref{sec:DataModel}), we must translate it. Figure~\ref{fig-xweb-data-tree-model} depicts the XWeB data tree model, which contains \emph{sale} facts, four dimensions (\emph{part}, \emph{customer}, \emph{supplier} and \emph{date}) and two measures (\textit{f\_quantity} and \textit{f\_totalamount}). The third row of Table~\ref{table-data-size} lists the sizes of the corresponding instances (minimum size is 27~MB and maximum size is 135~MB). 

Then again, XWeB's data warehouse does not include any complex hierarchy. Thus, we create variants of the dataset with different configurations of hierarchies: incomplete only, non-strict only and complex. 
Moreover, complexity is distributed by percentage of total number of dimensional nodes. For example, in Table~\ref{table-data-size}, which features the sizes of these datasets (last six rows), ``Complex 5\%'' on 50,000 facts means that among 200,000 dimensional nodes ($50,000 \times 4$ since each fact refers to four dimensions), 10,000 nodes are made complex. Such nodes are randomly distributed among every 20 ($100/5$) dimensional nodes. Moreover, the value of each generated node is randomly selected w.r.t. its dimensional applicable values, e.g., a \emph{month} node must contain numerical values between 1 and 12. Note that although Table~\ref{table-data-size} shows data sizes only for the 5\% and 50\% configurations, we also exploit intermediate configurations, i.e., 10\% and 20\%. 
In Table~\ref{table-data-size}, also note that data size expectingly decreases in incomplete configurations, since some subnodes are deleted, while data size increases in non-strict configurations, since subnodes are added to some dimensional nodes. Globally, data size increases in complex configurations since increases due to non-strictness are greater than decreases due to incompleteness.

Among XWeB's workload of queries, we focus on four queries with various number of dimensions (labeled 1D to 4D), and select the most detailed hierarchy levels for grouping because they form more complex groups. As shown in Table~\ref{table-group-by-dimensions}, we roll up to levels \emph{day}, \emph{type3}, \emph{nation} and \emph{nation} of dimensions \emph{date}, \emph{part}, \emph{customer} and \emph{supplier}, respectively. $n$ represents the number of dimension involved in a given query. The $sum$ aggregation function is used in our experiments to compute the total sale amount $sum(f\_totalamount)$. Any other aggregation function could be used, though.

\begin{table}[hbt]
\centering
\caption{Group by $n$-dimensions queries}
\label{table-group-by-dimensions}
{\small
\begin{tabular}{|l|l|l|l|l|} 
\hline
$n$&part&customer&supplier&date\\ 
\hline
1D&&&&day\\ \hline
2D&type3&&&day\\ \hline
3D&type3&nation&&day\\ \hline
4D&type3&nation&nation&day\\ \hline
\end{tabular}}
\end{table}

Finally, our experiments run on a Toshiba laptop with an Intel(R) Core(TM) i7-2670QM CPU @ 2.20~GHz, 4~GB memory and 64-bit Windows 7 Home Premium, Service Pack 1. The \texttt{QBS}, \texttt{MakeCovering} and \texttt{MakeStrict} algorithms are implemented in Java JDK 1.7, using the SAX parser to read XML data. The only difference between our Java code and Algorithm 1 is that the output tree is built on the fly instead of applying a product on intermediary trees, to optimize performance.

\subsection{Experimental Results}
\label{subsec-experimental-results}

The following subsections present the results of our experimental comparison of \texttt{QBS} and \texttt{Pedersen}. To perform this comparison, we created metadata so that \texttt{MakeCovering} replaces incomplete values by ``Other'' like \texttt{QBS} does. For \texttt{Pedersen}, we also differentiate between query execution time and preprocessing overhead, while we cannot for \texttt{QBS} since it operates at query time and overhead is confused with query execution time.

\subsubsection{Results on Simple Hierarchies}
\label{subsubsec-experimental-result-on-noncomplex-XWeb}

\begin{figure} [hbt]
	\centering
	\epsfig{file=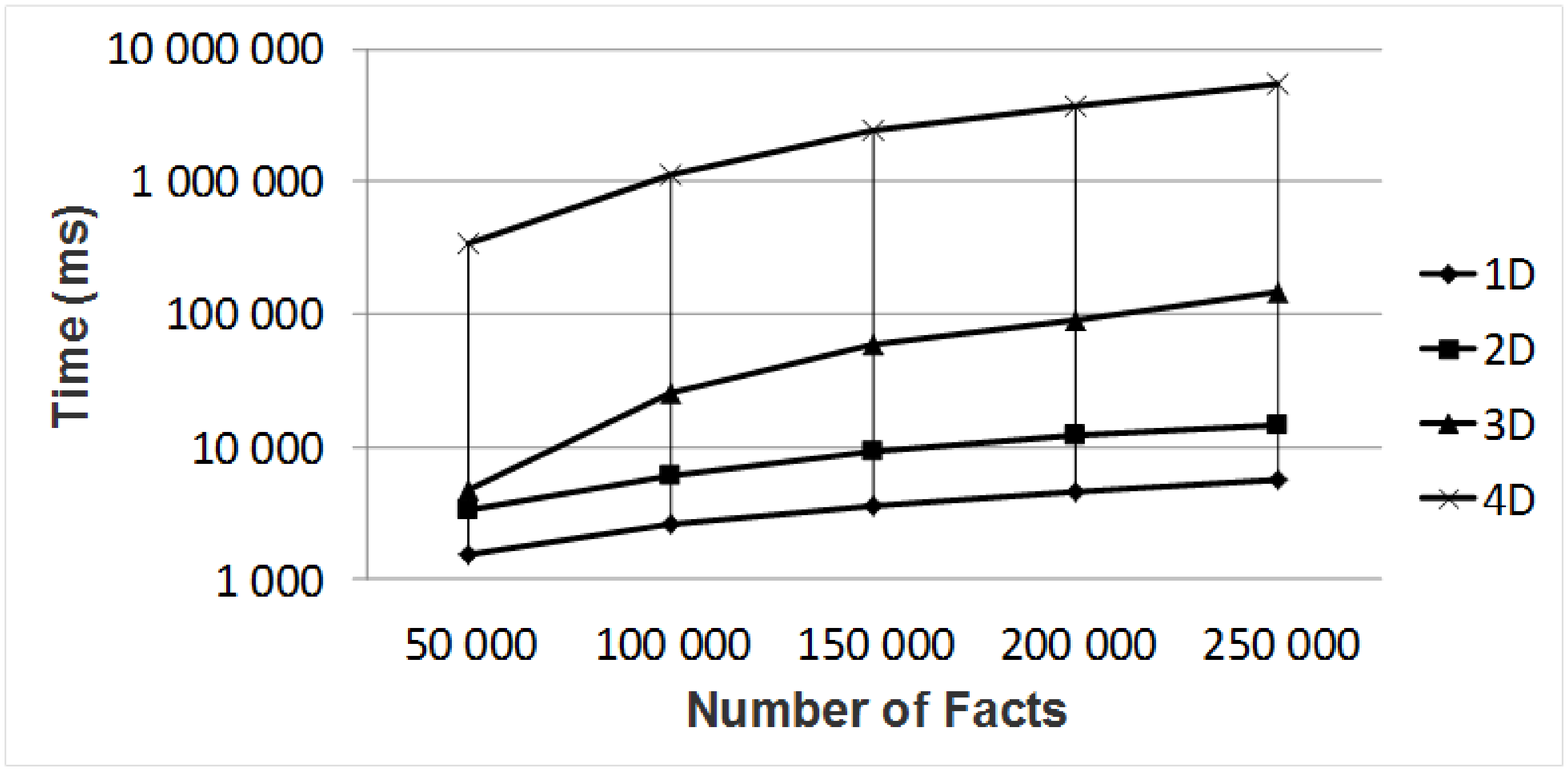, width = 8.3 cm}
	\caption{\texttt{QBS'} execution time according to the numbers of facts and of dimensions}
	\label{Fig12}
	\end{figure}

Figure~\ref{Fig12} shows that \texttt{QBS}' time performance increases linearly with data size (i.e., the number of facts) and the number of dimensions in the query, except for query 3D on 50,000 facts, which incidentally bears a lower grouping complexity. 

\begin{figure} [hbt]
	\centering
	\epsfig{file=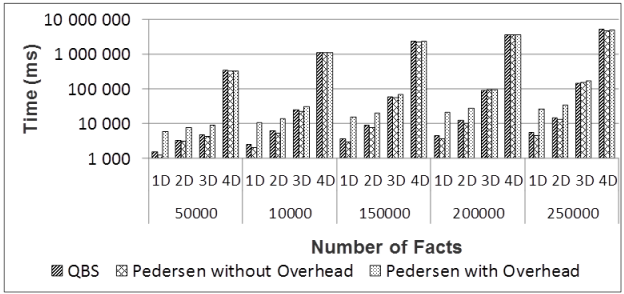, width = 8.5 cm}
	\caption{Comparison of \texttt{QBS} and \texttt{Pedersen} on simple hierarchies}  
	\label{Fig13}
	\end{figure}

Figure~\ref{Fig13} shows that the time performance of both approaches increases linearly w.r.t. data size and the number of dimensions used in queries. 

On average, the execution time of \texttt{QBS} is 2 times lower than that of \texttt{Pedersen with overhead}, but it is 0.17 times higher than that of \texttt{Pedersen without overhead}.

\begin{figure} [hbt]
	\centering
	\epsfig{file=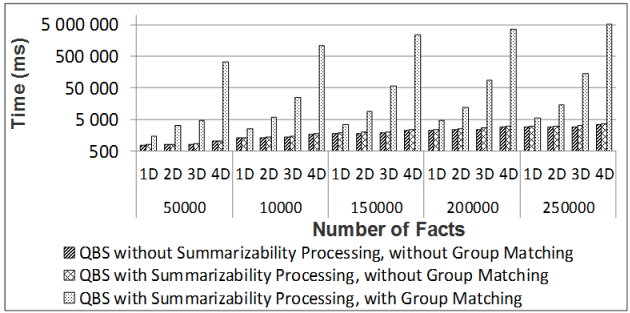, width = 8.5 cm}
	\caption{Comparison of summarizability processing time and group matching time in \texttt{QBS}} 
	\label{Fig14}
	\end{figure}

However, both \texttt{QBS} and \texttt{Pedersen} consume a lot of time, especially when running the 4D query (about an hour). To find out why, we perform two more experiments, dissociating complex hierarchy processing time (i.e., summarizability processing time) from group matching time. This is possible because XWeB's data are originally summarizable. Figure~\ref{Fig14} shows that enforcing summarizability in \texttt{QBS} does not affect time performance much, while group matching has a great impact that increases with the number of dimensions. 

\begin{figure} [hbt]
	\centering
	\epsfig{file=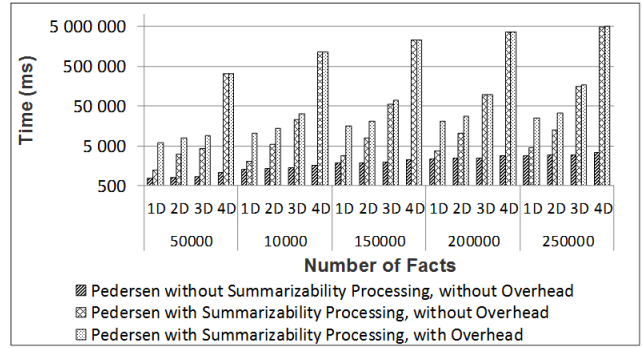, width = 8.5 cm}
	\caption{Comparison of summarizability processing time and group matching (overhead) time in \texttt{Pedersen}}
	\label{Fig15}
	\end{figure}

Figure~\ref{Fig15} confirms that \texttt{Pedersen} also spends most of its time processing group matching, while overhead consumes little time. When processing group matching, we indeed need to check whether the group exists.

Thus, we must check every hierarchy level instance in the whole group, which contains several instances from all dimensions. Doing so is very time consuming comparing to traditional aggregation, which only checks for the existing group as a whole. However, no approach dealing with XML grouping, and \emph{a fortiori} no XOLAP approach, can avoid this issue.



\subsubsection{Results on Complex Hierarchies}
\label{subsubsec-experimental-result-on-generated-complex-tpch}

Due to space limitations, we only present here our experiments on 5\% and 50\% incomplete, non-strict and complex hierarchies (the approximate minimum and maximum scale), but we did go through the whole range. \\
\\


\paragraph{Incomplete Hierarchies}
\label{sec:IncompleteHierarchies}

The results from Figures~\ref{Fig16} and \ref{Fig17} reveal two cases. When the number of dimensions is small (up to query 2D), {the execution time of \texttt{QBS} is 0.9 times lower than that of \texttt{Pedersen with overhead}, for both 5\% and 50\% hierarchies, on average. 

	\begin{figure} [hbt]
	\centering
	\epsfig{file=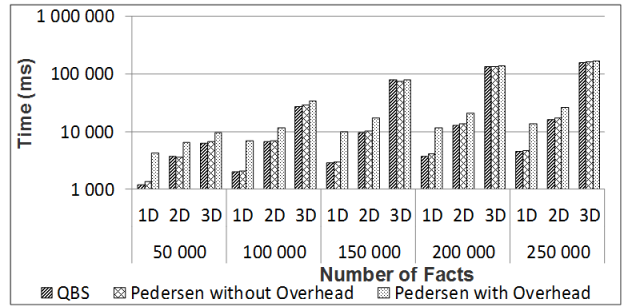, width = 8.5 cm}
	\caption{Comparison of \texttt{QBS} and \texttt{Pedersen} on 5\% incomplete hierarchies} 
	\label{Fig16}
	\end{figure}
	
	\begin{figure} [hbt]
	\centering
	\epsfig{file=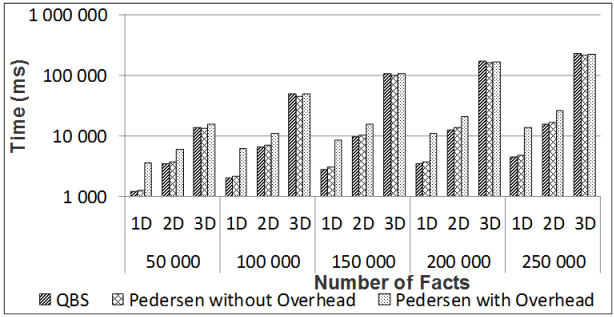, width = 8.5 cm}
	\caption{Comparison of \texttt{QBS} and \texttt{Pedersen} on 50\% incomplete hierarchies}
	\label{Fig17}
	\end{figure}

When overhead is not included in \texttt{Pedersen}, the execution time of \texttt{QBS} is 0.04 times lower (i.e., extremely close) on 5\% hierarchies and 0.02 times lower (i.e., extremely close) on 50\% hierarchies, on average}. For a larger number of dimensions (query 3D), {the execution time of \texttt{QBS} is the same as \texttt{Pedersen without overhead} on 5\% hierarchies and 0.06 times lower (i.e., extremely close) than that of \texttt{Pedersen without overhead} on 50\% hierarchies, on average. When overhead is included in \texttt{Pedersen}, \texttt{QBS'} execution time is on average 0.2 and 0.06 times lower (i.e., extremely close), on 5\% and 50\% hierarchies, respectively.} Both approaches actually have different tradeoffs. \texttt{QBS} takes less time when reading incomplete data, but more time to solve incompleteness, while the reverse is true for \texttt{Pedersen} where data are normalized. Thus, when the number of dimensions increases, \texttt{QBS}' overhead when processing incomplete hierarchies at run-time is a handicap that evens global performances w.r.t. \texttt{Pedersen}. Still, we can notice that both approaches are affected by the poor performance of group matching, which explains why we did not include query 4D in these experiments.\\
\\
\\

\paragraph{Non-Strict Hierarchies}
\label{sec:NonStrictHierarchies}

\begin{figure} [hbt]
	\centering
	\epsfig{file=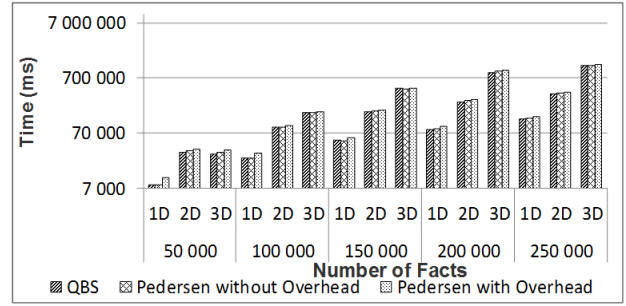, width = 8.5 cm}
	\caption{Comparison of \texttt{QBS} and \texttt{Pedersen} on 5\% non-strict hierarchies}
	\label{Fig18}
	\end{figure}

\begin{figure} [hbt]
	\centering
	\epsfig{file=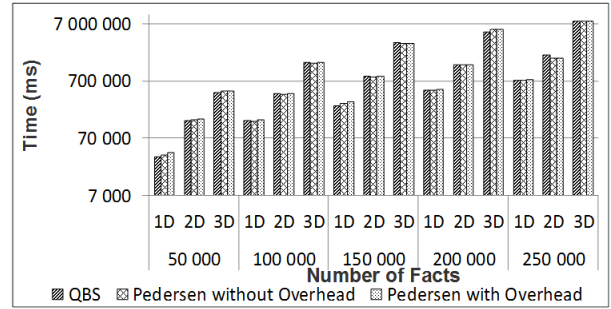, width = 8.5 cm}
	\caption{Comparison of \texttt{QBS} and \texttt{Pedersen} on 50\% non-strict hierarchies}
	\label{Fig19}
	\end{figure}

The results from Figures~\ref{Fig18} and \ref{Fig19} show similar trends to those of Figures~\ref{Fig16} and \ref{Fig17}, because the tradeoffs in \texttt{QBS} and \texttt{Pedersen} are essentially the same for non-strictness management.

\begin{figure} [hbt]
	\centering
	\epsfig{file=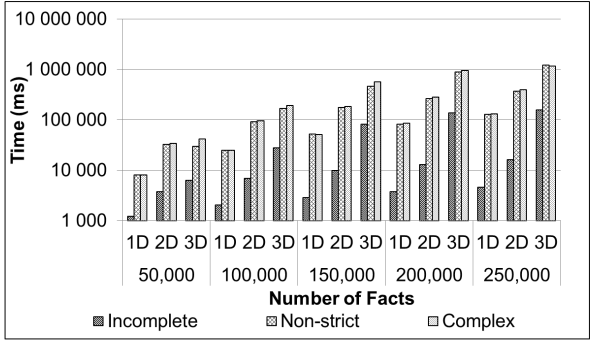, width = 8.5 cm}
	\caption{Evaluation of the three types of 5\% hierarchies in \texttt{QBS}}
	\label{Fig20}
	\end{figure}

\begin{figure} [hbt]
	\centering
	\epsfig{file=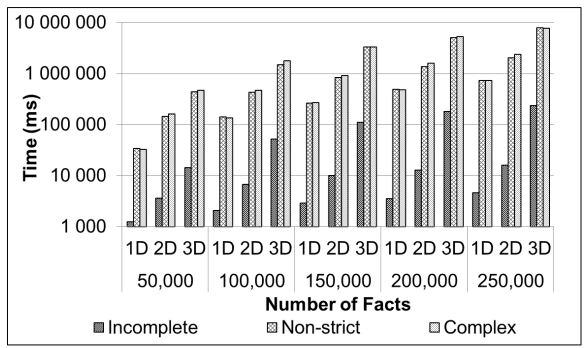, width = 8.5 cm}
	\caption{Evaluation of the three types of 50\% hierarchies in \texttt{QBS}}
	\label{Fig21}
	\end{figure}

However, for \texttt{QBS}, non-strictness processing is 9 times higher than incompleteness processing, on average (Figure~\ref{Fig20}). Moreover, non-strictness processing is 37 times higher than incompleteness processing, on average (Figure~\ref{Fig21}).

Ultimately, the execution time of \texttt{QBS} is 0.1 times lower than that of \texttt{Pedersen with overhead} (5\% hierarchies) and 0.03 times lower (i.e., extremely close) than that of \texttt{Pedersen with overhead}, on average (50\% hierarchies). When overhead is not included in \texttt{Pedersen}, the execution time of \texttt{QBS} is on average 0.05 times lower (5\% hierarchies) and 0.01 times lower (50\% hierarchies) (i.e., extremely close).


\paragraph{Complex Hierarchies}
\label{sec:ComplexHierarchies}

\begin{figure} [hbt]
	\centering
	\epsfig{file=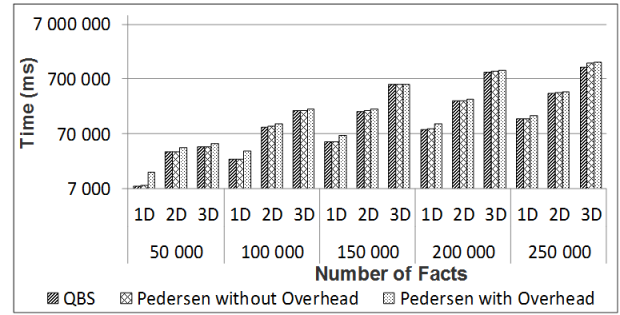, width = 8.5 cm}
	\caption{Comparison of \texttt{QBS} and \texttt{Pedersen} on 5\% complex hierarchies}
	\label{Fig22}
	\end{figure}

\begin{figure} [hbt]
	\centering
	\epsfig{file=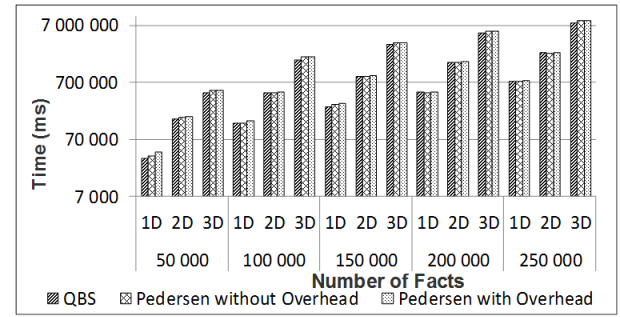, width = 8.5 cm}
	\caption{Comparison of \texttt{QBS} and \texttt{Pedersen} on 50\% complex hierarchies}
	\label{Fig23}
	\end{figure}

The results from Figures~\ref{Fig22} and \ref{Fig23} bear similar results to the non-strict case, again because the cost of non-strictness processing is much higher than that of incompleteness processing (Figures~\ref{Fig20} and \ref{Fig21}). 

		Group matching is indeed mainly impacted by non-strict hierarchies. However, in some cases, such as in the 3D query on 250,000 facts in Figure~\ref{Fig20}, \texttt{QBS} performs better in the complex case than in the non-strict case, because non-strict processing incidentally produces fewer complex groups, thus simplifying group matching. For 5\% hierarchies, \texttt{QBS'} execution time is 1.8 times lower than that of \texttt{Pedersen with overhead} and 0.01 times lower (i.e., extremely close) than that of \texttt{Pedersen without overhead}, on average. For 50\% hierarchies, \texttt{QBS'} execution time is 0.09 times lower (i.e., extremely close) than that of \texttt{Pedersen with overhead} and 0.05 lower (i.e., extremely close) than that of \texttt{Pedersen without overhead}, on average.\\
\\


\section{Conclusion and Perspectives}
\label{sec-conclusion}

In this paper, we propose the first truly operational query-based approach to solve summarizability issues in XML complex hierarchies. With respect to existing approaches, ours (1) modifies neither schema nor data, and thus has no space overhead and does not alter schema nor data semantics; (2)~does not require  any expertise beyond the user's, thus sparing the cost of expert intervention; (3) is dynamic w.r.t. schema and data evolution, thus favoring scalability. 

We indeed experimentally demonstrate that the overhead induced by managing hierarchy complexity at run-time is totally acceptable. The performance, in terms of query response time, of our \texttt{QBS} algorithm is indeed comparable to that of Pedersen et al.'s reference algorithms. However, our comparison holds when the dataset is static. If schema or data updates were made, complex hierarchy processing would take place at regular intervals of time with \texttt{Pedersen} (instead of once in our experiments). By contrast, \texttt{QBS} would not have any further overhead, and should thus become more efficient. 

Finally, our approach is implemented as a free Java prototype that is available online, along with our experimental datasets and the source code of the \texttt{QBS} and \texttt{Pedersen} algorithms\footnote{\url{http://eric.univ-lyon2.fr/~mhachicha/XOLAP.zip}}.

The perspectives of this work are twofold. First, although XML is the best-suited format to represent complex hierarchy structures, our experiments show that summarizability management approaches are still too costly for realistic OLAP processing, which is supposed to run \emph{online}, due to group matching cost. Thus, it is crucial to optimize the performance of our approach, e.g., by storing data in a non XML native fashion and/or using effective sorting, indexing and parallel processing techniques in group matching. 

In a second step, we aim to define other XOLAP operators (cube, drill down, etc.) over complex hierarchies in order to complete an algebra, and implement them in our software prototype to provide a fully operational XOLAP framework.

\bibliographystyle{abbrv}
\bibliography{mybib}

\begin{thebibliography}{10}

\bibitem{AbelloSS06}
A.~Abell{\'o}, J.~Samos, and F.~Saltor.
\newblock {YAM$^{\mbox{2}}$: a multidimensional conceptual model extending
  UML}.
\newblock {\em Information Systems}, 31(6):541--567, 2006.

\bibitem{BerglundBCFKRS10}
A.~Berglund, S.~Boag, D.~Chamberlin, M.~F. Fern{\'a}ndez, M.~Kay, J.~Robie, and
  J.~Sim{\'e}on.
\newblock {XML Path Language (XPath) 2.0 (Second Edition)}.
\newblock http://www.w3.org/TR/xpath20/, 2010.

\bibitem{BeyerCCOPX05}
K.~S. Beyer, D.~D. Chamberlin, L.~S. Colby, F.~{\"O}zcan, H.~Pirahesh, and
  Y.~Xu.
\newblock {Extending XQuery for Analytics}.
\newblock In {\em 24th International Conference on Management of Data (SIGMOD
  05), Baltimore, USA}, pages 503--514, 2005.

\bibitem{DyresonPJ03}
C.~E. Dyreson, T.~B. Pedersen, and C.~S. Jensen.
\newblock {Incomplete Information in Multidimensional Databases}.
\newblock In M.~Rafanelli, editor, {\em Multidimensional Databases: Problems
  and Solutions}, pages 282--309. Idea Group, 2003.

\bibitem{GolfarelliMR98}
M.~Golfarelli, D.~Maio, and S.~Rizzi.
\newblock {The Dimensional Fact Model: A Conceptual Model for Data Warehouses}.
\newblock {\em International Journal of Cooperative Information Systems},
  7(2-3):215--247, 1998.

\bibitem{HachichaD-tkde12}
M.~Hachicha and J.~Darmont.
\newblock {A Survey of XML Tree Patterns}.
\newblock {\em IEEE Transactions on Knowledge and Data Engineering}, 2012.
\newblock In preprint.

\bibitem{Hackathorn99}
R.~D. Hackathorn.
\newblock {\em Web farming for the data warehouse}.
\newblock The Morgan Kaufmann Series in Data Management Systems. Morgan
  Kaufmann, San Francisco, USA, 1999.

\bibitem{HornerS05}
J.~Horner and I.-Y. Song.
\newblock {A Taxonomy of Inaccurate Summaries and Their Management in OLAP
  Systems}.
\newblock In {\em 24th International Conference on Conceptual Modeling (ER 05),
  Klagenfurt, Austria}, volume 3716 of {\em LNCS}, pages 433--448. Springer,
  2005.

\bibitem{HummerLBS02}
W.~H{\"u}mmer, W.~Lehner, A.~Bauer, and L.~Schlesinger.
\newblock {A Decathlon in Multidimensional Modeling: Open Issues and Some
  Solutions}.
\newblock In {\em 4th International Conference on Data Warehousing and
  Knowledge Discovery (DaWaK 02), Aix-en-Provence, France}, volume 2454 of {\em
  LNCS}, pages 275--285. Springer, 2002.

\bibitem{HurtadoGM05}
C.~A. Hurtado, C.~Guti{\'e}rrez, and A.~O. Mendelzon.
\newblock {Capturing Summarizability with Integrity Constraints in OLAP}.
\newblock {\em ACM Transactions on Database Systems}, 30(3):854--886, 2005.

\bibitem{kimball02}
R.~Kimball and M.~Ross.
\newblock {\em The Data Warehouse Toolkit}.
\newblock John Wiley \& Sons, second edition, 2002.

\bibitem{LechtenborgerV03}
J.~Lechtenb{\"o}rger and G.~Vossen.
\newblock {Multidimensional Normal Forms for Data Warehouse Design}.
\newblock {\em Information Systems}, 28(5):415--434, 2003.

\bibitem{LenzS97}
H.-J. Lenz and A.~Shoshani.
\newblock {Summarizability in OLAP and Statistical Data Bases}.
\newblock In {\em 9th International Conference on Scientific and Statistical
  Database Management (SSDBM 97), Olympia, Washington, USA}, pages 132--143.
  IEEE Computer Society, 1997.

\bibitem{LiSZY05}
Z.~Li, J.~Sun, J.~Zhao, and H.~Yu.
\newblock {Transforming Non-covering Dimensions in OLAP}.
\newblock In {\em 7th Asia-Pacific Conference (APWeb 05), Shanghai, China},
  volume 3399 of {\em LNCS}, pages 381--393. Springer, 2005.

\bibitem{xweb-tpctc10}
H.~Mahboubi and J.~Darmont.
\newblock {XWeB: the XML Warehouse Benchmark}.
\newblock In {\em 2nd TPC Technology Conference on Performance Evaluation \&
  Benchmarking (TPCTC 10), Singapore}, volume 6417 of {\em LNCS}, pages
  185--203. Springer, September 2011.

\bibitem{MalinowskiZ06}
E.~Malinowski and E.~Zim{\'a}nyi.
\newblock Hierarchies in a multidimensional model: from conceptual modeling to
  logical representation.
\newblock {\em Data \& Knowledge Engineering}, 59(2):348--377, 2006.

\bibitem{Malinowski08z}
E.~Malinowski and E.~Zim{\'a}nyi.
\newblock {\em {Advanced Data Warehouse Design}}.
\newblock Springer, Berlin, Heidelberg, Germany, 2008.

\bibitem{MansmannS06}
S.~Mansmann and M.~H. Scholl.
\newblock {Extending Visual OLAP for Handling Irregular Dimensional
  Hierarchies}.
\newblock In {\em 8th International Conference on Data Warehousing and
  Knowledge Discovery (DaWaK 06), Krakow, Poland}, volume 4081 of {\em Lecture
  Notes in Computer Science}, pages 95--105. Springer, 2006.

\bibitem{MansmannS07}
S.~Mansmann and M.~H. Scholl.
\newblock {Empowering the OLAP Technology to Support Complex Dimension
  Hierarchies}.
\newblock {\em International Journal of Data Warehousing and Mining},
  3(4):31--50, 2007.

\bibitem{MazonLT08}
J.-N. Maz{\'o}n, J.~Lechtenb{\"o}rger, and J.~Trujillo.
\newblock {Solving Summarizability Problems in Fact-Dimension Relationships for
  Multidimensional Models}.
\newblock In {\em ACM 11th International Workshop on Data Warehousing and OLAP
  (DOLAP 08), Napa Valley, USA}, pages 57--64, 2008.

\bibitem{MazonLT09}
J.-N. Maz{\'o}n, J.~Lechtenb{\"o}rger, and J.~Trujillo.
\newblock {A survey on summarizability issues in multidimensional modeling}.
\newblock {\em Data {\&} Knowledge Engineering}, 68(12):1452--1469, 2009.

\bibitem{PedersenPP04}
D.~Pedersen, J.~Pedersen, and T.~B. Pedersen.
\newblock {Integrating XML Data in the TARGIT OLAP System}.
\newblock In {\em 20th International Conference on Data Engineering (ICDE 04),
  Boston, USA}, pages 778--781. IEEE Computer Society, 2004.

\bibitem{PedersenRP02b}
D.~Pedersen, K.~Riis, and T.~B. Pedersen.
\newblock {A Powerful and SQL-Compatible Data Model and Query Language for
  OLAP}.
\newblock In {\em 13th Australasian Database Conference (ADC 02), Melbourne,
  Australia}, volume~5 of {\em CRPIT}. Australian Computer Society, 2002.

\bibitem{PedersenJD99}
T.~B. Pedersen, C.~S. Jensen, and C.~E. Dyreson.
\newblock {Extending Practical Pre-Aggregation in On-Line Analytical
  Processing}.
\newblock In {\em 25th International Conference on Very Large Data Bases (VLDB
  99), Edinburgh, Scotland, UK}, pages 663--674. Morgan Kaufmann, 1999.

\bibitem{PourabbasR00}
E.~Pourabbas and M.~Rafanelli.
\newblock {Hierarchies and Relative Operators in the OLAP Environment}.
\newblock {\em SIGMOD Record}, 29(1):32--37, 2000.

\bibitem{RafanelliS90}
M.~Rafanelli and A.~Shoshani.
\newblock {STORM: A Statistical Object Representation Model}.
\newblock In {\em 5th International Conference on Statistical and Scientific
  Database Management (SSDBM 90), Charlotte, NC, USA}, volume 420 of {\em
  LNCS}. Springer, 1990.

\bibitem{Riz07}
S.~Rizzi.
\newblock {Conceptual Modeling Solutions for the Data Warehouse}.
\newblock In R.~Wrembel and E.~Christian~Koncilia, editors, {\em {Data
  Warehouses and OLAP: Concepts, Architectures and Solutions}}, pages 1--26.
  {IRM Press}, Hershey, USA, 2007.

\bibitem{RizziALT06}
S.~Rizzi, A.~Abell{\'o}, J.~Lechtenb{\"o}rger, and J.~Trujillo.
\newblock {Research in data warehouse modeling and design: dead or alive?}
\newblock In {\em ACM 9th International Workshop on Data Warehousing and OLAP
  (DOLAP 06), Arlington, Virginia, USA}, pages 3--10. ACM, 2006.

\bibitem{SongRME01}
I.-Y. Song, W.~Rowen, C.~Medsker, and E.~F. Ewen.
\newblock {An Analysis of Many-to-Many Relationships Between Fact and Dimension
  Tables in Dimensional Modeling}.
\newblock In {\em 3rd International Workshop on Design and Management of Data
  Warehouses (DMDW 01), Interlaken, Switzerland}, volume~39 of {\em CEUR
  Workshop Proceedings}, page~6. CEUR-WS.org, 2001.

\bibitem{Tor03}
R.~Torlone.
\newblock {Conceptual Multidimensional Models}.
\newblock In E.~Maurizio~Rafanelli, editor, {\em {Multidimensional Databases:
  Problems and Solutions}}, pages 69--90. {IDEA Group Publishing}, Hershey,
  USA, 2003.

\bibitem{tpch}
{TPC}.
\newblock {\em {TPC Benchmark H Standard Specification revision 2.3.0}}.
\newblock Transaction Processing Performance Council, August 2005.

\end{thebibliography}

\end{document}